\documentclass{article}

\usepackage{arxiv}

\usepackage[utf8]{inputenc} 
\usepackage[T1]{fontenc}    
\usepackage{hyperref}       
\usepackage{url}            
\usepackage{booktabs}       
\usepackage{amsfonts}       
\usepackage{nicefrac}       
\usepackage{microtype}      
\usepackage{graphicx}
\usepackage[square,numbers]{natbib}
\usepackage{amssymb}
\usepackage{amsmath}
\usepackage{doi}

\title{Noise and fluctuations can undermine the efficiency of Majority Rule in Group Evaluation problems}

\author{ \href{https://orcid.org/0000-0002-3485-9249}{\includegraphics[scale=0.06]{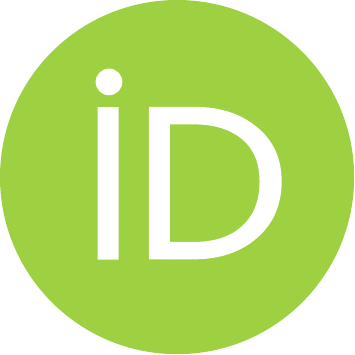}\hspace{1mm}Daniele Vilone} \\ 
	Laboratory of Agent Based Social Simulation, Institute of Cognitive Science and Technology,\\
	National Research Council, Via Palestro 32, 00185 Rome, Italy;\\
	\ \ \\
	Grupo Interdisciplinar de Sistemas Complejos, Departamento de Matem\'aticas,\\
	Universidad Carlos III de Madrid, 28911 Legan\'es, Spain\\
	\ \ \\
	(\texttt{daniele.vilone@gmail.com}) \\
}

\date{}

\renewcommand{\shorttitle}{Noise and fluctuations in Group Evaluation problems}

\begin{document}
\maketitle

\begin{abstract}
	Crowdsourcing is a mechanism by means of which groups of people are able to execute a task by sharing ideas, efforts and resources. Thanks to the online technologies, crowdsourcing has become in the last decade an even more utilized process in different and diverse fields. An instance of such process is the so-called "label aggregation problem": in practice, it is the evaluation of an item by groups of agents, where each agent gives its own judgment of it. Starting from the individual evaluations of their members, how can the groups give their global assessment? In this work, by means of a game-theoretical, evolutionary approach, we show that in most cases the majority rule (the group evaluation is the evaluation of the majority of its members) is still the best way to get a reliable group evaluation, even when the agents are not the best experts of the topic at stake; on the other hand, we also show that noise ({\it i.e.}, fortuitous errors, misunderstanding, or every possible source of non-deterministic outcomes) can undermine the efficiency of the procedure in non-trivial situations. Therefore, in order to make the process as reliable as possible, the presence of noise and its effects should be carefully taken into account.
\end{abstract}


\section{Introduction}
\label{intro} 

Crowdsourcing is a process in which a given problem is solved or a challenge addressed by the effort of a mass of people~\cite{brabham2013crowdsourcing,ghezzi2018crowdsourcing}. An immediate case of crowdsourcing is fund-raising, but in recent years we have seen relevant instances in different fields, in particular scientific research~\cite{mclaughlin2014nasa}. 
The emergence and even more widespread use of social networks has further widened the range of possible applications of crowdsourcing, because now a very large number of people may be involved to accomplish a given task by the new online technologies. Together with other factors, this has also paved the way to the so-called Citizen Science~\cite{gura2013citizen}, by means of which also common people, but interested in science, has helped professional scientists to get new results and/or solve problems.

Therefore, in order to understand its basic mechanisms, learn how to manage and use it correctly, and improve its efficiency~\cite{li2016crowdsourcing,brabham2012crowdsourcing}, crowdsourcing itself has become a hot research topic. More in general, crowdsourcing can be viewed as a particular case of cooperative behaviour, so that it can be studied also by means of Game Theory and evolutionary dynamics~\cite{guazzini2015modeling}.

In this work, by means of a game-theoretical approach, we will present an evolutionary simulation model aimed to understand the dynamics and main features of a peculiar case of Citizen Science, which has heavily attracted the interest of many researchers~\cite{cerquides2021conceptual,brabham2012myth}. In the next sections we will first describe the problem, then define the model and study it, finally we will draw some conclusions and possible perspectives for future works.

\section{The label aggregation problem}
\label{label_agg} 

We start by considering the situation where a group of experts have made their evaluation on a given topic, and on the basis of such judgements a final evaluation has to be drawn~\cite{collins2009latent}. For example, a classical instance is when many clinicians have given a diagnosis and an overall evaluation is required to make a medical decision~\cite{dawid1979maximum}. In general, these issues used to involve experts on the field, so that there is arguably a limited margin of error. On the other hand, today we have similar situations where experts do not participate alone in evaluation processes~\cite{inel2014crowdtruth}: indeed, today Citizen Science and crowdsourcing can be also accomplished by involving non-professional volunteers in social networks~\cite{brabham2012myth,imran2014aidr}. In particular, in this work we focus on the case of a natural disaster and the need to evaluate the entity of damages only from pictures taken {\it in loco}~\cite{cerquides2021conceptual}: a photo depicting a damage is sent to a group of people who give their assessment, and eventually a final, conclusive group evaluation has to be delivered. The research question we want to address here is how can we process the individual assessments in order to get the most reliable group evaluation. For this purpose we are going to utilize the tool of evolutionary simulations: within a game-theoretical framework, we set a population where agents with different behaviours/features act, and make them undergo a process of natural selection so that the most efficient traits finally emerge~\cite{roca2009evolutionary}. In this respect, we stress the fact that our evolutionary model is not meant to describe how some observable traits originated in humans, but only to compare the efficiency of each trait with respect to the others.

More in depth, the process we are to study here works as follows (see Ref.~\cite{cerquides2021conceptual}). A picture is given to a group of $N$ people. Each member of the group has to evaluate the level of damages reported in the picture by assigning to it a note which is an integer from 0 to 4 (0: irrelevant; 1: no-damage; 2: minimal; 3: moderate; 4: severe). Therefore, based on the individuals' assessments, the group's evaluation is given: this is the core of the label aggregation problem. The elementary process just described (evaluating a picture as a group) is called {\it task}. In general,  for each event we have many groups which accomplish many similar tasks. Consequently, it is crucial to find the best algorithm able to elaborate the individual evaluations and obtain the best group response, {\it i.e.}, able to minimize the number of wrong group evaluations.

The most immediate choice for the algorithm is the {\it majority rule} (MR): for each task, the note given by the majority of the group members is the one adopted by the group as a whole (for instance, if we have a group with 5 members, and in a specific task three of them have evaluated the picture as a 3, 3 will be the group evaluation), if more than one note are the most voted, one of them is selected at random. Interestingly, the majority rule has already been studied as a tool for getting a decision in groups in different contexts, in particular to analyze the emergence of leadership~\cite{galam1990social}, the spreading of minority opinions~\cite{galam2002minority}, and the effect of contrarians~\cite{galam2004contrarian}.

Beyond MR, other algorithms are possible, in particular non-deterministic ones~\cite{cerquides2021conceptual}. Indeed, adding noise in the dynamics shows to improve the efficiency of a process in different situations, as for example reaching consensus in a population~\cite{sen2014sociophysics} or to increase the level of cooperation in evolving populations~\cite{hofbauer1998evolutionary,nowak2004evolutionary}: it is worth then to check the effect of noise also in the collective evaluation of an item. The effect of noise is not universal, because from one hand a non-deterministic rule increases the errors produced during the process, but on the other one it makes the system explore a larger portion of the configuration space~\cite{bottcher2021computational}. Therefore, the details of the algorithm and the system under scrutiny are of fundamental importance to evaluate the efficiency of a non-deterministic algorithm with respect to the classical MR.
In this work we will compare the outcome of the majority rule\footnote{MR is {\it per se} a deterministic rule but, as we illustrate in Sec.~\ref{themodel}, noise intervenes in the individual evaluations of the group members.}  with a noisy one, which we will call {\it weighted random rule} (WR) and define precisely in the next section. Next to MR and WR, in order to have a baseline, we will also consider the {\it totally random rule} (RR), where groups evaluate items completely at random.

\section{The Model}
\label{themodel} 

In this section we define a general model to simulate the processes involved in the label aggregation problem. Our aim is to extract the main features of the model and single out the best options to make it the most efficient possible. We will investigate the behaviour of the model as a function of different parameters, evaluation rules and evolutionary algorithms.

\subsection{The Algorithm of the dynamics}
\label{algor} 

We consider a population of $N$ individuals distributed in $M$ groups of $G$ agents each, (obviously, it must hold $N=M\times G$). There are three kinds of agents: inexperienced, volunteers (which are not totally inexperienced but far from being professionals) and experts, and every group is made up by agents of the same type (only experts, only volunteers, or only inexperienced individuals). 
At each elementary step of the dynamics, one group is picked up at random and has to accomplish a task, that is, the group must evaluate a generic item: there are $l$ categories of items $\{0,1,\dots,l-1\}$, and the group's aim is to guess the correct category of the item. For each task, the group gains fitness according to its evaluation. Subsequently, another group is randomly selected and accomplishes a task, and so on. One time unit is given by $M$ elementary steps (Monte Carlo steps). Every $T$ time units there is a reproduction stage. This model is in principle more general than the one of Reference~\cite{cerquides2021conceptual}, where $l=5$.

\ 

\noindent {\bf Initial conditions --} All the groups have initial fitness equal to zero. The initial distribution of the agent types is set in order to be as realistic as possible: therefore, we set a $5\%$ of groups composed by experts, $15\%$ of volunteers, and the remaining $80\%$ of inexperienced agents.

\ 

\noindent {\bf Evaluation process: agents --} When a group must accomplish a task, its members evaluate the category $c$ of the item. The value $c$ is an integer between $0$ and $l-1$, and it is assigned at random. Each agent evaluates the item according to its own kind: inexperienced guess the right category of the item with a probability $p_{ine}$, volunteers with probability $p_{vol}$, experts with probability $p_{exp}$. It is naturally $p_{ine}<p_{vol}<p_{exp}$; if they do not guess the right answer, they will then assign to the item a random number between 0 and $l$ but different from $c$ itself.

\noindent In this work we have set $p_{ine}=0.20$, $p_{vol}=0.50$ and $p_{exp}=0.95$.

\ 

\noindent {\bf Evaluation process: groups --} Once every member has given its evaluation, the group as a whole has to give its final assessment. There are three possible ways for a group to accomplish the task: guessing completely at random (baseline, totally random rule), letting a randomly selected member to give its evaluation for the group (weighted random rule), or according to the majority rule (already introduced in the previous section).

\ 

\noindent {\bf Fitness updating --} After the group has given its evaluation, if the task has been correctly accomplished (that is, the group has ascertained the right category), its fitness increases by 1 unit, otherwise it remains unchanged.

\ 

\noindent {\bf Asynchronous dynamics --} We stress the fact that this dynamics is asynchronous, as usual in Monte Carlo simulations~\cite{newman1999monte}, that is, every time a task is assigned, the group is extracted at random, so that during each time unit every group solves one task only on average. As already suggested in Sec.~\ref{label_agg}, this procedure introduces noise that allows the system to explore a larger part of the configuration space also with MR.

\ 

\noindent {\bf Evolution stage --} After every $T$ Monte Carlo steps of the dynamics, the evolution stage takes place. This is instead a synchronous process: every group picks up another one at random and compare the fitness: if the second group has a better fitness, the first one imitates the latter, that is, its members become of the same kind of the better performing group ({\it i.e.}, if the extracted group has higher fitness and it is made up by experts, also the members of the first one become experts if they are not), then the fitness is set to zero again. In order to maintain heterogeneity in the dynamics, in this work we have assumed the value $T=1$, considering the case $T>1$ just to check the role of noise over the model dynamics (see Subsec.~\ref{mesosc}).
This evolutionary dynamics is necessary not because some real, physical processes similar to the label aggregation problem actually evolve, but in order to compare the efficiency of the three types of agents: given a method to get the group evaluation (MR, WR, or RR), which one is the best performing kind of agents? The ability to survive in a competitive dynamics shows how effective a given strategy is.

\ 

\noindent {\bf Fixed agent type, evaluation algorithms evolving --} We also accomplished a bunch of simulations where instead of setting the group evaluation algorithm, the agent type is fixed for all: in that case, the groups throughout the system differentiate not for the kind of their members (experts, volunteers, inexperienced), but for the process adopted to get the group evaluation (MR, WR, RR). Therefore, in these simulations the initial conditions are of course different: we start with $1/3$ of groups adopting MR, $1/3$ WR, and $1/3$ RR with all the agents of the same kind.

\ 

\noindent {\bf Ensemble averages --} Every simulation for each set of parameter values chosen is repeated a minimum of 1000 up to 20000 times, all the results shown are averages overs such number of independent realizations ({\it ensemble averages})~\cite{newman1999monte,huang1963statistical}. 

\section{Results}
\label{results}

In this section we are presenting the main results of our simulations. The model is quite general, so that we have a bunch of parameters to set: group size $G$, number of groups $M$, number of possible evaluation values $l$. In order to get a systematic description of the behaviour of the model, first we set $l=5$ as in Ref.~\cite{cerquides2021conceptual}, then we split the discussion by considering separately the small group case (in particular $G=5$) and the limit of very large $G$.

\subsection{Small groups ($G=5$)}

\begin{figure}
  \centering
  \includegraphics[width=113mm]{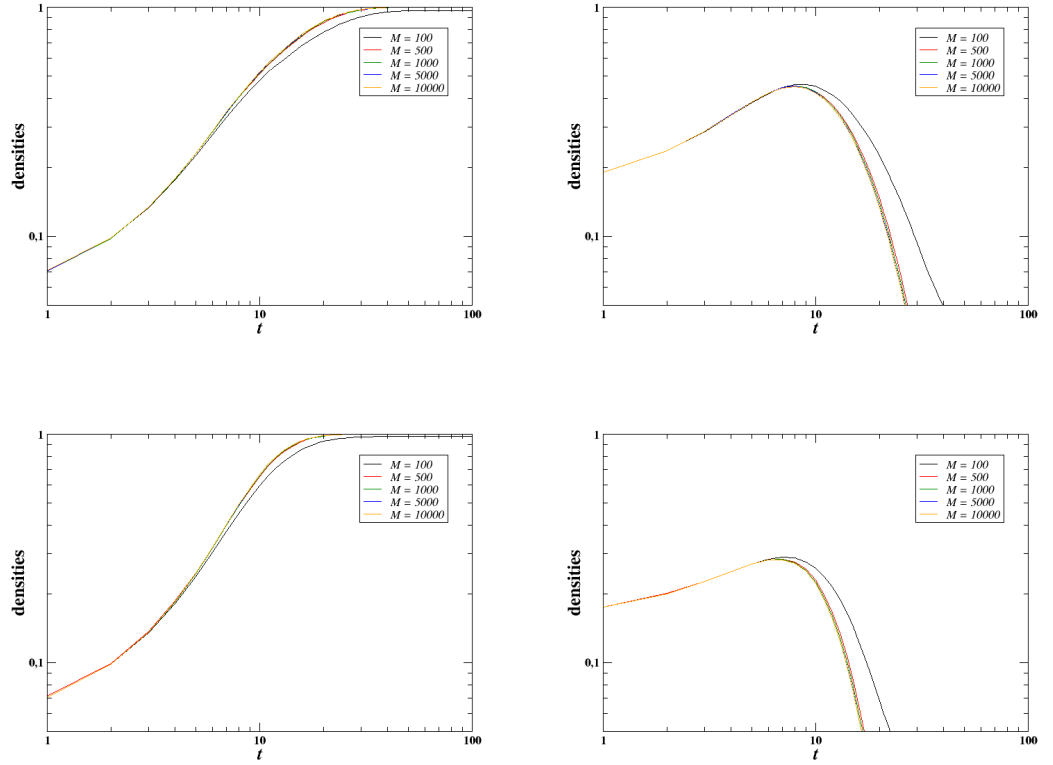}
  \caption{Expert (left column) and volunteers (right column) densities as functions of time, for different system sizes ($M$: number of groups), $G=5$, $l=5$; top row: Majority rule; bottom row: Weighted random rule.}
  \label{smallG_01}
\end{figure}

Let us consider the case $G=5$, as in Ref.~\cite{cerquides2021conceptual}.
In Fig.~\ref{smallG_01} (top row) we observe the time behaviour of the expert and volunteer densities as functions of time for different system sizes (since it is $N=M\times G$, we selected $M$ as control parameter for the whole population size) when the group evaluation is established by means of the MR. As it is easy to see, for $M\gtrsim500$ the system appears to have reached the thermodynamic limit. It is clear that MR makes experts far more efficient to accomplish the task than less experienced individuals, and the same holds with WR, as shown in the same Fig.~\ref{smallG_01} (bottom row). Inexperienced agents are very weak, and in every parameter configuration they disappear quickly from the system.

\begin{figure}
  \centering
  \includegraphics[width=131mm]{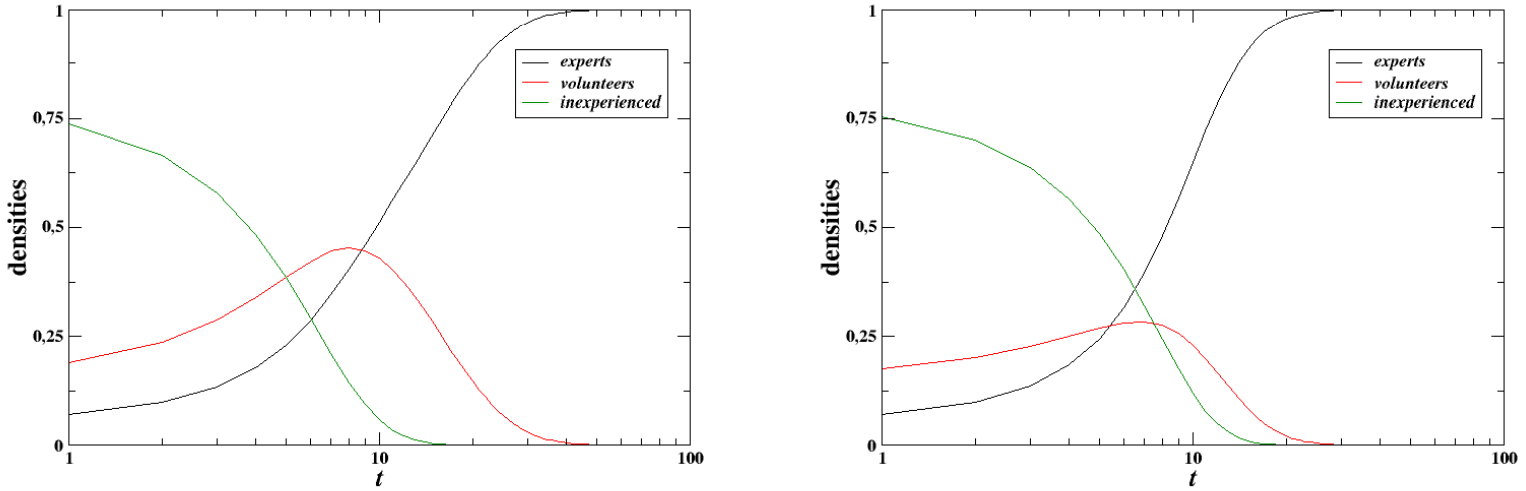}
  \caption{{\bf Left:} Time behaviour of the densities of the three agent types with majority rule to accomplish group evaluation. {\bf Right:} Time behaviour of the densities of the three agent types with weighted random rule to accomplish group evaluation. For both graphs the remaining parameter values are $M=500$, $G=5$, $l=5$.}
  \label{smallG_02}
\end{figure}

In short, by one side experts are always the best subjects to form an evaluation group (as it was easy to expect); by another one, if a group is made up by volunteers with some knowledge about the issue, it is more efficient when the majority rule is adopted as the procedure to decide. Indeed, volunteers resist better and longer with MR with respect to WR, as it is clear from Fig.~\ref{smallG_02}: whilst the evolutionary dynamics makes them vanish against experts in both cases, we remind that the goal of such simulations is to evaluate the efficiency of the different types of agents and decision algorithms. 

Anyway, in order to get more reliable results in this respect, we prepared some simulations in which all the agents are of the same kind, but different groups with different methods to reach their evaluation (MR, WR and RR) initially coexist: therefore, here the decision algorithms compete instead of agent types. In the following paragraph we illustrate these results.

\subsubsection*{Evolving evaluation algorithms}

\begin{figure}
  \centering
  \includegraphics[width=131mm]{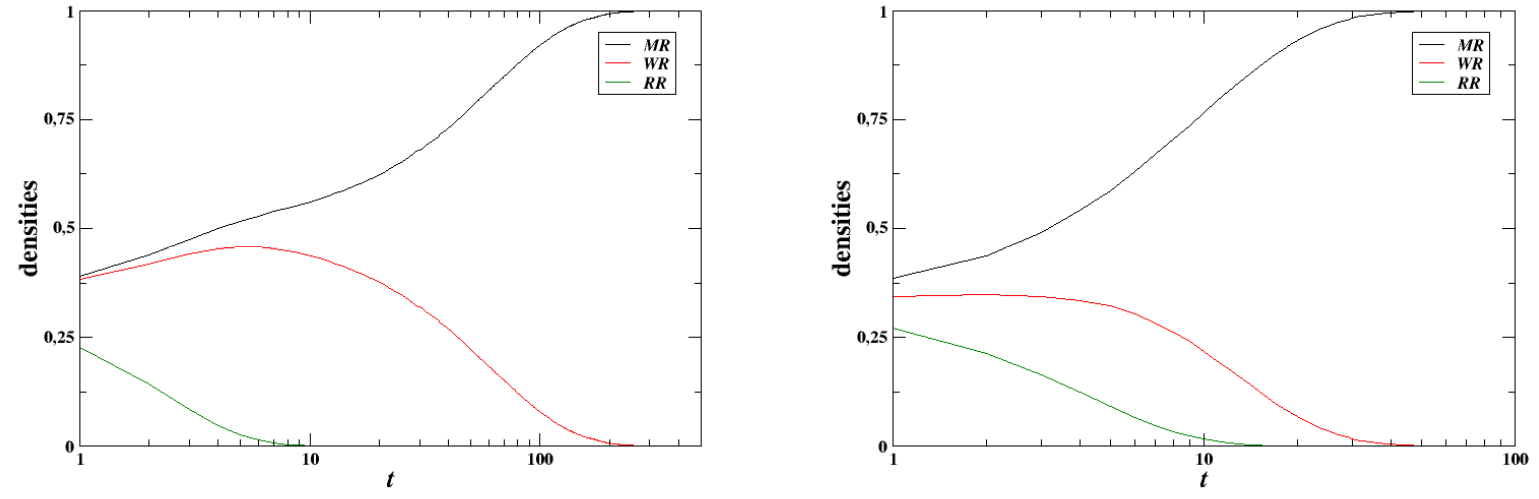}  \caption{Time behaviour of the densities of groups adopting majority rule (black lines), weighted random rule (red lines) and random rule (green lines). {\bf Left:} all agents are experts; {\bf right:} all agents are volunteers. The remaining parameter values are $M=500$, $G=5$, $l=5$.}
  \label{algor_compet}
\end{figure}

In Figure~\ref{algor_compet} we show the behaviour of a system made up by only experts (left) or volunteers (right), where groups distinguish themselves for the evaluation algorithm: in both cases we see that MR is the more efficient method, and RR is (comprehensibly) the worst one. On the other hand, a very different behaviour is observed in a population of inexperienced agents: in this case, the three group evaluation rules are equivalent, indeed the (average) densities of the three possible algorithms do not change during the dynamics and maintain their initial value, that is, 1/3 each one (we do not show the graphics because it is trivial). Therefore, when agents are completely non-experts, no evaluation rule guarantees reliable group decisions.

\subsection{Large groups ($G\gg5$)}

\begin{figure}
  \centering
  \includegraphics[width=113mm]{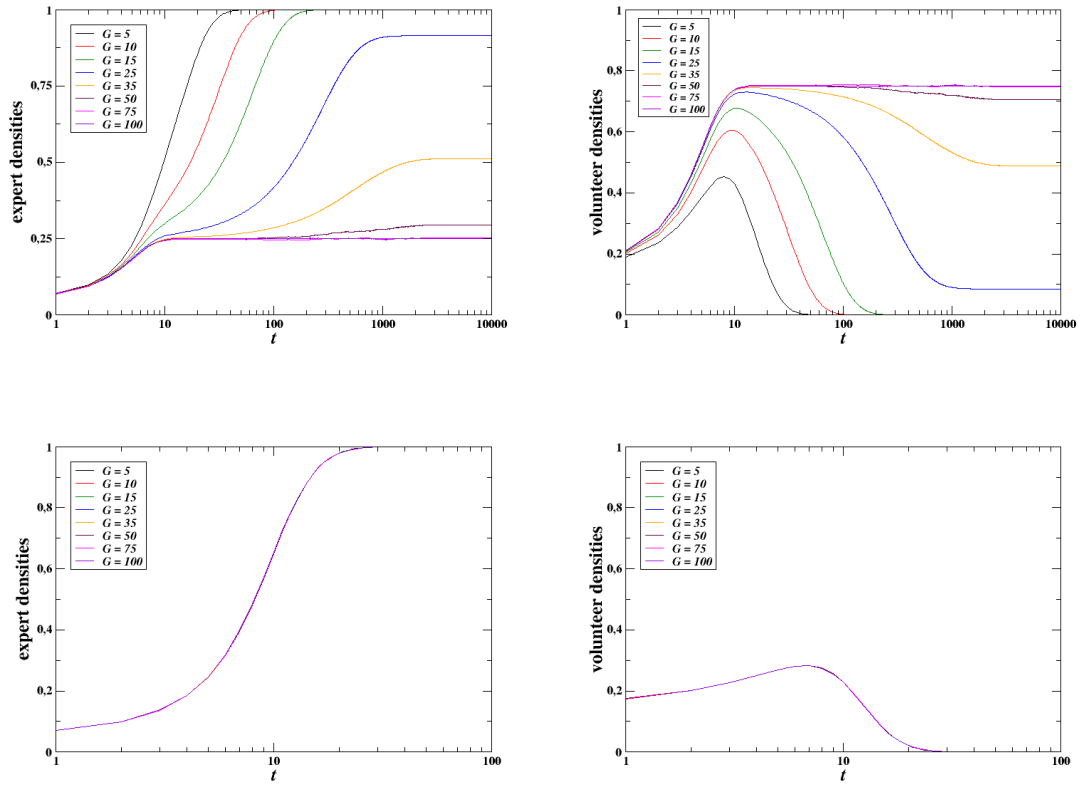}
  \caption{Expert (left column) and volunteer (right column) densities as functions of time, for different group sizes $G$, $M=500$, $l=5$; top row: Majority rule; bottom row: Weighted random rule. With WR the parameter $G$ has no practical effect, so that all the different curves overlap.}
  \label{largeG_01}
\end{figure}

So far, we have dealt with groups composed by few members ($G=5$). What happens if we consider larger groups? In Figure~\ref{largeG_01} we show the behaviour of the expert (left) and volunteer (right) densities for a system where agent types evolve, with majority (above) and weighted random (below) rule, $M=500$ and $l=5$, and several values of the parameter $G$. The first thing worth to be noticed is the deep difference among MR and WR dynamics. Indeed, with WR rule the size of the groups has no influence, while $G$ has a fundamental role with MR.

In particular, if the group size is large enough, with the majority rule experts do not invade the hole system, but it ends up to a final, mixed configuration where experts and volunteers coexist (inexperienced density always quickly vanishes in every case). It results particularly interesting the fact that for some large systems volunteers, without invading completely the system, end up being more than experts, that is, at least in such conditions volunteers act more efficiently than experts.

\begin{figure}
  \centering
  \includegraphics[width=83mm]{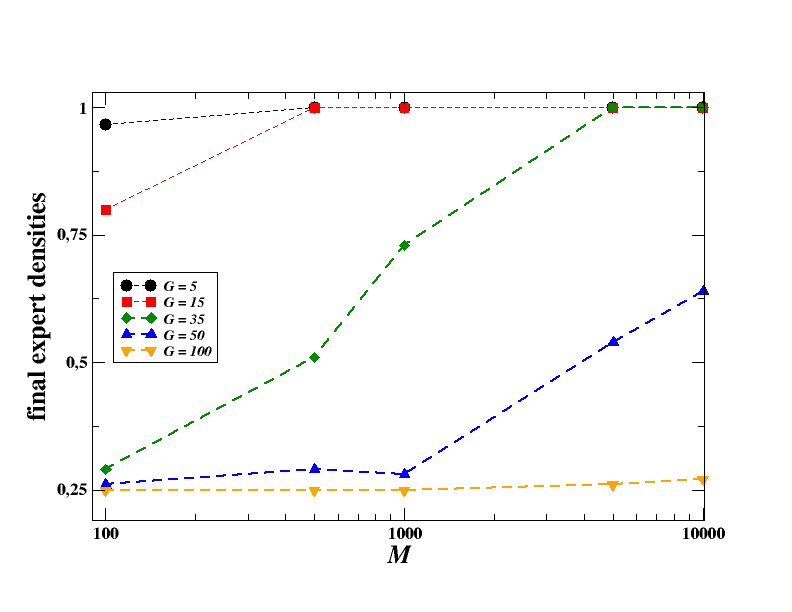}
  \caption{Behaviour of the final expert density as a function of the system size for a system with majority rule, $l=5$, and several values of $G$. The correspondent final values of the volunteer densities is simply given by $1$ minus the expert final density (inexperienced agents always disappear).}
  \label{largeG_02}
\end{figure}

On the other hand, from Figure~\ref{largeG_02} it seems that in the thermodynamic limit ($M\rightarrow+\infty$) expert density increases for every value of $G$. Unfortunately, for numerical reasons it is very hard to investigate systematically what happens at the end of the dynamics in very large systems, as it happens to be for high values of both $G$ and $M$, so that we can not conclude definitively that in the thermodynamic limit experts invade the population again (even though it is very likely, see Sec.~\ref{discuss}), but it is already noticeable that for many values of $M$ and $G$ majority rule appears to privilege volunteers instead of experts.

\section{Discussion}
\label{discuss}

In general, our simulations show two main results. The first, predictable one is that experts are more efficient than volunteers, on their turn more efficient than inexperienced agents. Secondly, that majority rule is the best group evaluation algorithm in most cases. This behaviour can be explained in terms of probability to correctly evaluate a task. More precisely, with the weighted random rule the probability $P_x^W$ (being $x$ the agent type {\it exp, vol} or {\it ine}) for a group to catch the right evaluation is simply the probability to solve the task for the single agent:

\begin{equation}
    P_x^W=p_x \ , 
    \label{probWR}
\end{equation}

\noindent while with the majority rule this probability $P_x^M$ has got a more complex expression:

\begin{equation}
P_x^M=\Gamma_x + \Pi_{res}^x \ .
\label{probMR}
\end{equation}

\noindent In Eq.~(\ref{probMR}) $\Gamma_x$ is the probability that the right answer has been given by the absolute majority of group members, and $\Pi_{res}^x$ the residual probability to give the right group evaluation even though only a minority of members gave the right answer (see Subsec.~\ref{algor}); $x$ indicates again the agent type. The first term at right-hand side of Eq.~(\ref{probMR}) can be easily written down:

\begin{equation}
    \Gamma_x = \sum_{k=K_G}^G\binom{G}{k}\ p_x^k(1-p_x)^{G-k} \ , 
    \label{Gamma_def}
\end{equation}

\noindent where $K_G=1+[G/2]$ (being $[Z]$ the floor of $Z$). 

\begin{figure}
  \centering
  \includegraphics[width=83mm]{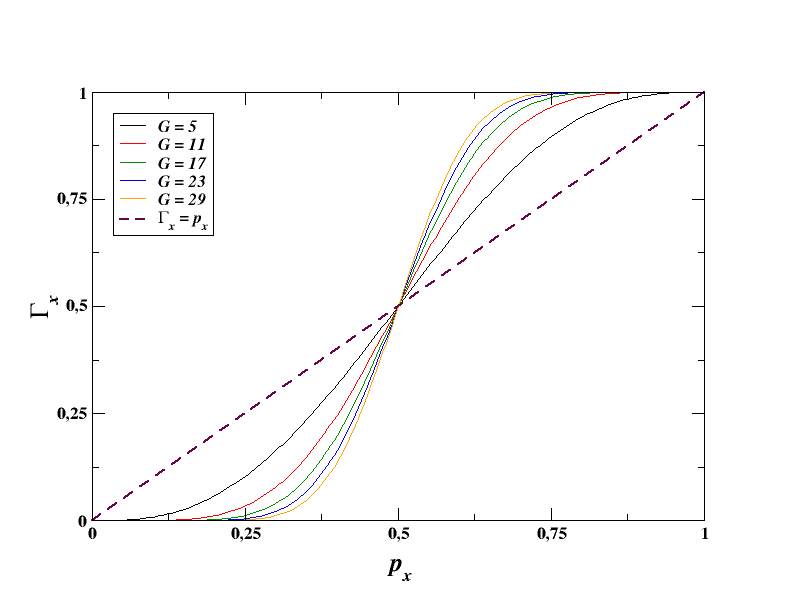}
  \caption{Behaviour of the function $\Gamma_x(p_x)$, as defined in Eq.~(\ref{Gamma_def}), for different values of the parameter $G$. The maroon, dashed line practically shows the probability $P_x^W$ as function of $p_x$ with WR as defined in Eq.~(\ref{probWR}). See also Fig. 2 in Ref.~\cite{galam1990social}. }
  \label{sigmoids}
\end{figure}

As it may be easily computed numerically, the function $\Gamma_x$ defined in Eq.~(\ref{Gamma_def}) is a sigmoid in its argument (see Fig.~\ref{sigmoids}); a deeper study of its mathematical features can be found in Ref.~\cite{galam1990social}.

The consequences of these results are the following. First of all, both $P_x^W$ and $\Gamma_x$ are strictly increasing functions of their arguments, that is, the higher is the value of the elementary probability $p_x$, the higher is the probability for a group to solve a task. Secondly, for $p_x>1/2$, $\Gamma_x$ is always larger than $P_x^W$, so that, following Eq.~(\ref{probMR}), it is always $P_x^M>P_x^W$, meaning that, once fixed the parameters of the model, MR is generally more efficient than WR. Of course, RR is the worst decision algorithm, being in this case $P_x^R=1/G\ \forall x$.

The numerical results shown in Sec.~\ref{results} agree with this conclusions always for small values of $G$, always with WR algorithm, and in the limit $M\rightarrow+\infty$ with MR (see Fig.~\ref{largeG_02}). More complex appears the behaviour of the model with majority rule, large $G$ and intermediate values of the system size, as we are going to describe in the next subsection.

\subsection{\label{mesosc}Mesoscale behaviour: the role of noise}

As illustrated in Sec.~\ref{results}, there is a set of values of the parameters $M$ and $G$ such that with the majority rule, volunteers outperform experts in an evolutionary framework. More in depth, we see that after a non-trivial transient, expert and volunteer densities reach a plateau at a value different from 0 and 1 (inexperienced agents get always extinct quickly). Therefore, in these circumstances experts and volunteers coexist, and for $G$ large enough volunteers remain the majority. On the other hand, from what we found previously, by computing the probabilities of all the possible outcomes, the experts should be always favourite against volunteers: in fact, such probabilistic calculations neglect finite size effects, that is, ignore any possible influence of noise and fluctuations on the dynamics. Therefore, for system sizes not too big with respect to the groups' size, the presence of noise and fluctuations, due to the Monte Carlo dynamics, is realistically the source of this counterintuitive outcome. We can verify the plausibility of this conclusion by decreasing the influence of noise on the dynamics. In order to do that, we performed some simulations increasing the value of $T$, that is, increasing the the time in between the evolution stages: up to now, we have maintained $T=1$, which means that each group plays on average once before evolving and setting again its fitness equal to zero. Indeed, with $T=1$ every evolutionary stage will take place with some groups which have played more than once while others never have, so that, for example, we can have some volunteer groups that have solved the task (getting a fitness larger than zero) and expert groups which remained with zero fitness. Then, by increasing the value of $T$ we let every group to play more than once before each evolution stage, and this protects expert groups from the risk to be outperformed by volunteers.

As a matter of fact, we considered the case with $G=35$ of Fig.~\ref{largeG_01} with MR (remaining parameters: $l=5,\ M=500$). With the ordinary value $T=1$ we see that the final volunteer density is around 0.47 (experts at $\sim0.53$); similar simulations with $T=5$ show a time behaviour qualitatively very similar (the maximum of volunteer density is the same, for instance), but the final volunteer density stays around 0.27; analogously, with $T=10$ the final volunteer density is about 0.14. It is then plausible to infer that in the limit of very large $T$, when the effect of noise is completely removed, the final configuration of the system will be expert density equal to 1 and volunteers extinct, as suggested by previous theoretical calculations.

\section{Conclusions and perspectives}

 Following Ref.~\cite{cerquides2021conceptual}, in this work we have studied the so-called label aggregation problem. We considered groups made up by $G$ members which have to solve tasks consisting in evaluating a given item that in its turn can assume one of $l$ possible values. Every member of each group gives an evaluation, then the group as a whole, based on the components' answers, will give its global assessment. Such task can be coped with through different procedures and methods: group members can be more or less expert about the topic and groups can adopt different methods to elaborate the individual evaluations. We accomplished the investigation by means of a game-theoretical approach, that is, through evolutionary simulations, where groups and agents of different kind and utilizing different evaluation rules compete in order to make the most efficient procedure emerge.

In short, we have established by one side that the more expert are the agents, the more efficient is the evaluation by their groups, which is trivial, but also that the majority rule is more efficient that a simple probabilistic algorithm which does not discern among individuals' ability. On the other hand, the majority rule shows to be quite sensitive to noise for a non-trivial subset of the model parameters, in particular when the groups are quite large while the system as a whole is not too large. This, if confirmed by further studies, may represent a useful warning for users of this tool for crowdsourcing: a simple majority rule, even though theoretically very efficacious with respect to other procedures, can provide misleading responses when the groups are large but not numerous, and, more in general, if every possible source of noise is not suitably managed. Future works, be them theoretical, empirical, and/or simulations, should clarify this point.





\section*{Funding}
This work was partially supported by the project Humane-AI-net, from the Horizon 2020 Research and Innovation program of the European Union, grant agreement No. 820437.

\section*{Acknowledgments}
The author wants to thank Jesus Cerquides for the opportunity to discuss this model, and his useful suggestions about it.

\bibliographystyle{unsrtnat}
\bibliography{vilo2022arx}  

\begin{thebibliography}{23}
\providecommand{\natexlab}[1]{#1}
\providecommand{\url}[1]{\texttt{#1}}
\expandafter\ifx\csname urlstyle\endcsname\relax
  \providecommand{\doi}[1]{doi: #1}\else
  \providecommand{\doi}{doi: \begingroup \urlstyle{rm}\Url}\fi

\bibitem[Brabham(2013)]{brabham2013crowdsourcing}
Daren~C Brabham.
\newblock \emph{Crowdsourcing}.
\newblock Mit Press, 2013.

\bibitem[Ghezzi et~al.(2018)Ghezzi, Gabelloni, Martini, and
  Natalicchio]{ghezzi2018crowdsourcing}
Antonio Ghezzi, Donata Gabelloni, Antonella Martini, and Angelo Natalicchio.
\newblock Crowdsourcing: a review and suggestions for future research.
\newblock \emph{International Journal of Management Reviews}, 20\penalty0
  (2):\penalty0 343--363, 2018.

\bibitem[McLaughlin(2014)]{mclaughlin2014nasa}
Elliot~C. McLaughlin.
\newblock Image overload: Help us sort it all out, nasa requests,\newline
  https://edition.cnn.com/2014/08/17/tech/nasa-earth-images-help-needed.
\newblock
  \emph{https://edition.cnn.com/2014/08/17/tech/nasa-earth-images-help-needed/},
  2014.

\bibitem[Gura(2013)]{gura2013citizen}
Trisha Gura.
\newblock Citizen science: amateur experts.
\newblock \emph{Nature}, 496\penalty0 (7444):\penalty0 259--261, 2013.

\bibitem[Li et~al.(2016)Li, Huhns, Tsai, and Wu]{li2016crowdsourcing}
Wei Li, Michael~N Huhns, Wei-Tek Tsai, and Wenjun Wu.
\newblock \emph{CROWDSOURCING.}
\newblock Springer, 2016.

\bibitem[Brabham(2012{\natexlab{a}})]{brabham2012crowdsourcing}
Daren~C Brabham.
\newblock Crowdsourcing: A model for leveraging online communities.
\newblock In \emph{The participatory cultures handbook}, pages 138--147.
  Routledge, 2012{\natexlab{a}}.

\bibitem[Guazzini et~al.(2015)Guazzini, Vilone, Donati, Nardi, and
  Levnaji{\'c}]{guazzini2015modeling}
Andrea Guazzini, Daniele Vilone, Camillo Donati, Annalisa Nardi, and Zoran
  Levnaji{\'c}.
\newblock Modeling crowdsourcing as collective problem solving.
\newblock \emph{Scientific reports}, 5\penalty0 (1):\penalty0 1--11, 2015.

\bibitem[Cerquides et~al.(2021)Cerquides, M{\"u}l{\^a}yim,
  Hern{\'a}ndez-Gonz{\'a}lez, Ravi~Shankar, and
  Fernandez-Marquez]{cerquides2021conceptual}
Jesus Cerquides, Mehmet~O{\u{g}}uz M{\"u}l{\^a}yim, Jer{\'o}nimo
  Hern{\'a}ndez-Gonz{\'a}lez, Amudha Ravi~Shankar, and Jose~Luis
  Fernandez-Marquez.
\newblock A conceptual probabilistic framework for annotation aggregation of
  citizen science data.
\newblock \emph{Mathematics}, 9\penalty0 (8):\penalty0 875, 2021.

\bibitem[Brabham(2012{\natexlab{b}})]{brabham2012myth}
Daren~C Brabham.
\newblock The myth of amateur crowds: A critical discourse analysis of
  crowdsourcing coverage.
\newblock \emph{Information, Communication \& Society}, 15\penalty0
  (3):\penalty0 394--410, 2012{\natexlab{b}}.

\bibitem[Collins and Lanza(2009)]{collins2009latent}
Linda~M Collins and Stephanie~T Lanza.
\newblock \emph{Latent class and latent transition analysis: With applications
  in the social, behavioral, and health sciences}, volume 718.
\newblock John Wiley \& Sons, 2009.

\bibitem[Dawid and Skene(1979)]{dawid1979maximum}
Alexander~Philip Dawid and Allan~M Skene.
\newblock Maximum likelihood estimation of observer error-rates using the em
  algorithm.
\newblock \emph{Journal of the Royal Statistical Society: Series C (Applied
  Statistics)}, 28\penalty0 (1):\penalty0 20--28, 1979.

\bibitem[Inel et~al.(2014)Inel, Khamkham, Cristea, Dumitrache, Rutjes, van~der
  Ploeg, Romaszko, Aroyo, and Sips]{inel2014crowdtruth}
Oana Inel, Khalid Khamkham, Tatiana Cristea, Anca Dumitrache, Arne Rutjes,
  Jelle van~der Ploeg, Lukasz Romaszko, Lora Aroyo, and Robert-Jan Sips.
\newblock Crowdtruth: Machine-human computation framework for harnessing
  disagreement in gathering annotated data.
\newblock In \emph{International semantic web conference}, pages 486--504.
  Springer, 2014.

\bibitem[Imran et~al.(2014)Imran, Castillo, Lucas, Meier, and
  Vieweg]{imran2014aidr}
Muhammad Imran, Carlos Castillo, Ji~Lucas, Patrick Meier, and Sarah Vieweg.
\newblock Aidr: Artificial intelligence for disaster response.
\newblock In \emph{Proceedings of the 23rd international conference on world
  wide web}, pages 159--162, 2014.

\bibitem[Roca et~al.(2009)Roca, Cuesta, and S{\'a}nchez]{roca2009evolutionary}
Carlos~P Roca, Jos{\'e}~A Cuesta, and Angel S{\'a}nchez.
\newblock Evolutionary game theory: Temporal and spatial effects beyond
  replicator dynamics.
\newblock \emph{Physics of life reviews}, 6\penalty0 (4):\penalty0 208--249,
  2009.

\bibitem[Galam(1990)]{galam1990social}
Serge Galam.
\newblock Social paradoxes of majority rule voting and renormalization group.
\newblock \emph{Journal of Statistical Physics}, 61\penalty0 (3):\penalty0
  943--951, 1990.

\bibitem[Galam(2002)]{galam2002minority}
Serge Galam.
\newblock Minority opinion spreading in random geometry.
\newblock \emph{The European Physical Journal B-Condensed Matter and Complex
  Systems}, 25\penalty0 (4):\penalty0 403--406, 2002.

\bibitem[Galam(2004)]{galam2004contrarian}
Serge Galam.
\newblock Contrarian deterministic effects on opinion dynamics:“the hung
  elections scenario”.
\newblock \emph{Physica A: Statistical Mechanics and its Applications},
  333:\penalty0 453--460, 2004.

\bibitem[Sen and Chakrabarti(2014)]{sen2014sociophysics}
Parongama Sen and Bikas~K Chakrabarti.
\newblock \emph{Sociophysics: an introduction}.
\newblock Oxford University Press, 2014.

\bibitem[Hofbauer et~al.(1998)Hofbauer, Sigmund,
  et~al.]{hofbauer1998evolutionary}
Josef Hofbauer, Karl Sigmund, et~al.
\newblock \emph{Evolutionary games and population dynamics}.
\newblock Cambridge university press, 1998.

\bibitem[Nowak and Sigmund(2004)]{nowak2004evolutionary}
Martin~A Nowak and Karl Sigmund.
\newblock Evolutionary dynamics of biological games.
\newblock \emph{science}, 303\penalty0 (5659):\penalty0 793--799, 2004.

\bibitem[B{\"o}ttcher and Herrmann(2021)]{bottcher2021computational}
Lucas B{\"o}ttcher and Hans~J Herrmann.
\newblock \emph{Computational Statistical Physics}.
\newblock Cambridge University Press, 2021.

\bibitem[Newman and Barkema(1999)]{newman1999monte}
Mark~EJ Newman and Gerard~T Barkema.
\newblock \emph{Monte Carlo methods in statistical physics}.
\newblock Clarendon Press, 1999.

\bibitem[Huang(1963)]{huang1963statistical}
Kerson Huang.
\newblock Statistical mechanics, john wily \& sons.
\newblock \emph{New York}, page~10, 1963.

\end{thebibliography}






\end{document}